# Non-Fermi liquid normal state of the Heavy Fermion superconductor UBe$_{13}$

P. Gegenwart [a,*], C. Langhammer [a], R. Helfrich [a], N. Oeschler [a], M. Lang [b], J.S. Kim [c], G.R. Stewart [c], and F. Steglich [a]

[a] *Max-Planck Institute for Chemical Physics of Solids, D-01187 Dresden, Germany*

[b] *Institute of Physics, University of Frankfurt/Main, D-60054 Frankfurt, Germany*

[c] *Department of Physics, University of Florida, Gainesville, FL 32611, USA*

**Abstract**

Non-Fermi liquid (NFL) behavior in the normal state of the heavy-fermion superconductor UBe$_{13}$ is studied by means of low-temperature measurements of the specific heat, $C$, and electrical resistivity, $\rho$, on a high-quality single crystal in magnetic fields up to 15.5 T. At $B=0$, unconventional superconductivity forms at $T_c=0.9$ K out of an incoherent state, characterized by a large and strongly temperature dependent $\rho(T)$. In the magnetic field interval 4 T $\leq B \leq$ 10 T, $\rho(T)$ follows a $T^{3/2}$ behavior for $T_c(B) \leq T \leq 1$ K, while $\rho$ is proportional to $T$ at higher temperatures. Corresponding Non-Fermi liquid behavior is observed in $C/T$ as well and hints at a nearby antiferromagnetic (AF) quantum critical point (QCP) covered by the superconducting state. We speculate that the suppression of short-range AF correlations observed by thermal expansion and specific heat measurements below $T_L \approx 0.7$ K ($B=0$) yields a field-induced QCP, $T_L \to 0$, at $B=4.5$ T.

Non-Fermi liquid behavior, Heavy Fermion superconductivity, UBe$_{13}$

## 1. Introduction

Unconventional superconductivity has been the subject of intense research in recent decades. Of particular interest are those systems in which superconductivity develops out of a normal state that cannot be described within the Landau Fermi liquid model. Most prominent examples include the high-$T_c$ cuprates [1] and the heavy-fermion (HF) superconductors [2]. For some clean Ce-based HF systems which are antiferromagnets at ambient pressure, HF superconductivity has been discovered in a small pressure range around the quantum-critical point (QCP), where $T_N \to 0$, to develop out of a non-Fermi liquid (NFL) state, characterized by an unusual resistivity behavior $\rho - \rho_0 \propto T^\epsilon$ with $1 \leq \epsilon \leq 1.5$ and a strongly $T$-dependent specific heat coefficient $C/T$ [3,4]. This led to the proposal that in these systems superconductivity is mediated by an electronic rather than the usual elastic Cooper pairing interaction [3].

The cubic UBe$_{13}$ is one of the most fascinating HF superconductors [5] because i) superconductivity develops out of a highly unusual normal state characterized by a large and strongly $T$-dependent electrical resistivity [5,6] and ii) upon substituting a small amount of Th for U in U$_{1-x}$Th$_x$Be$_{13}$, a non-monotonic evolution of $T_c$ and a second phase transition at $T_{c2}$ below $T_{c1}$, the superconducting one, is observed in a critical concentration range $x_{c1}=0.019 < x < 0.045 = x_{c2}$ [7].

In this paper, we investigate the low-temperature normal-state NFL behavior of a high-quality single crystal of UBe$_{13}$ also studied in [8,9]. The $T$-dependences observed in specific heat (section 2) and electrical resistivity (section 3) resemble those found for Ce-based HF systems near the antiferromagnetic (AF) QCP. This leads to the speculation that the NFL behavior in UBe$_{13}$ is caused by a QCP at a finite magnetic field slightly above 4 T, covered by the superconducting (SC) phase. Indeed careful studies of the thermal expansion and specific heat have revealed a "line of anomalies", $B^*(T)$, presumably of short-range AF origin, in the $B$-$T$ phase diagram, that starts at $T_L=0.7$ K ($B=0$) and terminates at $B^*(0) \approx 4.5$ T [9]. These anomalies have been shown to mark the precursor of the lower lying of the two second-order phase transitions in the critical range $x_{c1} < x < x_{c2}$ in U$_{1-x}$Th$_x$Be$_{13}$ [9]. The $B$-$T$ phase diagram presented in section 4 suggests a relation between the suppression of the $T_L$ anomaly in UBe$_{13}$ within the SC state at 4.5 T and the normal-state NFL behavior.

*Corresponding author. Tel.: +49-351-4646-2324; fax: +49-351-4646-2360; e-mail: gegenwart@cpfs.mpg.de (Philipp Gegenwart).

## 2. Specific heat

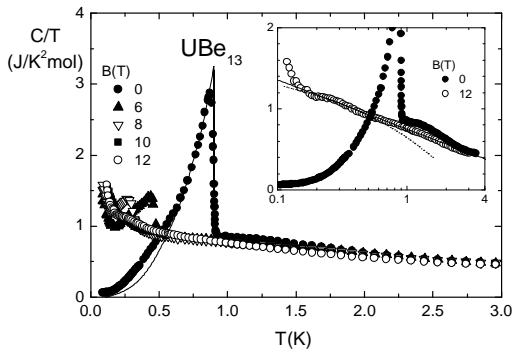

Fig. 1: Specific heat $C$ of a UBe$_{13}$ single crystal after subtraction from raw data the nuclear contribution due to Zeeman splitting of $^9$Be states [10], as $C/T$ vs $T$ at differing magnetic fields. Solid line indicates the quasiparticle contribution assuming an axial SC order parameter [10]. Inset shows 0 and 12 T data on a logarithmic temperature scale. Solid and dotted lines indicate $C/T \propto -\log T$ and $C/T = \gamma_0 - \beta T^{1/2}$, respectively.

The specific heat has been measured with the aid of a thermal-relaxation method at magnetic fields up to 12 T [10]. In Figure 1, the electronic specific heat coefficient is displayed between 60 mK and 3 K. The high quality of the single crystal is reflected in the sharpness of the phase transition anomaly at $T_c = 0.9$ K (10% to 90% width: 25 mK). The equal areas construction [9] reveals a large reduced jump $\Delta C^{sc}/\gamma_n T_c$ of 2.6, when using $\gamma_n=0.9$ J/K$^2$mol as the electronic specific heat coefficient at $T_c$, that would indicate a strong-coupling SC state. However, as can be seen from Figure 1, in order to balance the entropy between the normal and SC states, the normal-state specific heat coefficient has to increase towards $T \rightarrow 0$. A very similar observation has been made for the recently discovered Ce-based HF superconductor CeCoIn$_5$ [11]. The application of magnetic fields which suppress superconductivity enables the study of the normal state specific heat coefficient $C/T$ down to lower temperatures. Indeed an increase of $C/T$ towards $(1.4 \pm 0.2)$ J/K$^2$mol is observed at $B=12$ T (inset Figure 1. Here, the low-$T$ upturn in the 12 T data below 0.2 K has been ignored. The origin of the latter is unclear, since the expected Schottky contribution due to the Zeeman splitting of the nuclear $^9$Be nuclear spin states has already been subtracted from the raw data [10]. The origin of the normal state logarithmic increase of $C/T$, registered at $B=12$ T and resembling that observed for CeCoIn$_5$ at $B=5$ T ($\approx B_{c2}$) [11], will be discussed below.

## 3. Electrical resistivity

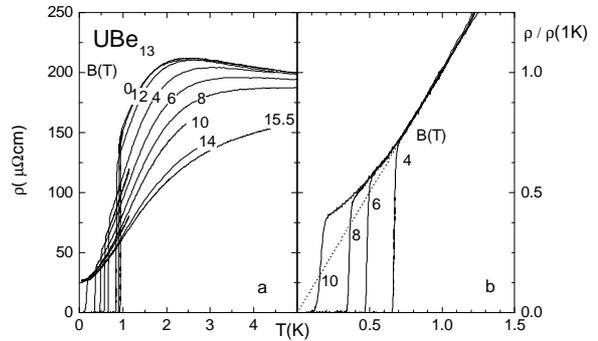

Fig. 2a: Electrical resistivity $\rho$ vs $T$ of a UBe$_{13}$ single crystal at $B=0$ and differing fields; (b): the same data as in (a), normalized to the respective $\rho(T)$ value at $T=1$ K. Dashed line is an extrapolation to $T=0$ of the data for $T \geq 0.8$ K.

The electrical resistivity of a small piece from the same high-quality single crystal studied in specific heat has been measured by utilizing the standard four-terminal ac-technique [12]. As displayed in Figure 2a, the zero-field resistivity $\rho(T)$ displays a pronounced maximum near $T_{max} \approx 2$ K. From the maximum value of the resistivity an inelastic mean free path as short as a few Å can be inferred, indicative of an incoherent metallic state, that remarkably shows a rather low magnetic susceptibility. This had led Cox [6] to propose a quadrupolar Kondo scenario for UBe$_{13}$ by assuming a 4+ (5$f^2$) Uranium valence state with a low-lying non-magnetic $\Gamma_3$ crystal-field (CF) doublet. This scenario reveals an *intrinsic* finite residual resistivity (at $B=0$) and a large negative magnetoresistance slope d$\rho$/d$B$ [13], as indeed observed, cf. Figure 2a. On the other hand, CF effects studied via specific heat [14] and Raman-scattering [15] experiments as well as measurements of the non-linear susceptibility [16] seem to support a trivalent (5$f^3$) configuration. The resistivity maximum at 2 K indicates a low-energy scale in addition to the characteristic "Kondo-lattice" scale, $T^*$, accounting for the extremely large effective carrier masses. It has been estimated that $T^*$ ranges from 8 K [14] to 30 K [17]. This additional energy scale manifests itself in a distinct maximum in the thermal expansion coefficient [18] and a less pronounced shoulder in specific heat around 2 K and may be ascribed to local spin fluctuations in disordered Kondo systems [9]. These "2K fluctuations" dominate the normal state resistivity for fields below 4 T. At larger fields, 4 T $\leq B \leq$ 10 T, we are able to scale the various $\rho(T)$ curves within $T_c(B) \leq T \leq 1.2$ K to a universal curve by normalizing $\rho(T)$ to its respective value at 1 K (Figure 2b). Above $T \approx 0.7$ K, the normal-state resistivity is found to be *proportional* to $T$, similar to what has been observed for optimally doped high-$T_c$ cuprates [19]. At lower temperatures, $\rho(T)$ becomes superlinear.

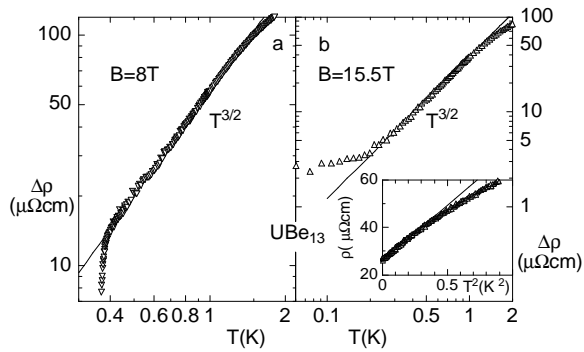

Fig. 3: Temperature dependent contribution $\Delta\rho = \rho - \rho_0$ of the electrical resistivity of a UBe$_{13}$ single crystal *vs* $T$ (on logarithmic scales) at $B=8$T (a) and $B=15.5$T (b), respectively. Inset shows low-$T$ data of (b) as $\rho$ *vs* $T^2$.

As demonstrated in Figure 3a for the $B = 8$ T data, $\Delta\rho(T) \propto T^{3/2}$ for $T_c(8\,\mathrm{T}) \leq T \leq 1$ K. Apparently, this $T$ dependence is in full accord with the theoretical prediction for a three-dimensional system of itinerant AF spin fluctuations in the vicinity of a QCP [20,21]. At elevated temperatures, the same scenario predicts a cross-over to a quasi-linear $T$-dependence of $\rho(T)$, as indeed observed. As shown in the inset of Figure 1, the specific heat coefficient follows $C/T \propto -\log T$ for $T \geq 0.3$ K and gradually deviates to smaller values at lower $T$ before the yet unexplained upturn sets in at the lowest temperatures. In the limited $T$ range 0.15 K $\leq T \leq 0.4$ K, the specific heat coefficient can be satisfactorily described by $C/T = \gamma_0 - \beta T^{1/2}$ which is expected along with the $T^{3/2}$ behavior of $\Delta\rho(T)$ [20,21]. Thus, for 4 T $\leq B \leq$ 12 T both electrical resistivity and specific heat show NFL behavior consistent with the nearness of an AF QCP, assuming three-dimensional critical AF fluctuations [20,21] and a substantial degree of potential scattering [22].

Since magnetic fields act on the critical spin-fluctuations associated with a magnetic QCP, a field-induced change from NFL to FL behavior is expected in these systems, as observed, e.g., at $B_0 \approx 6$ T in "S-type" CeCu$_2$Si$_2$ [23], $B_0 \approx 5$ T in CeCoIn$_5$ [24] and $B_0 \approx 1$ T in CeNi$_2$Ge$_2$ [25]. For UBe$_{13}$, this cross-over field is as large as 14 T at which $\Delta\rho(T) = A(B)T^2$ becomes visible below 0.3 K with a gigantic coefficient $A$. The latter is decreasing with increasing $B$ from 52 $\mu\Omega$cm/K$^2$ at 14 T to 45 $\mu\Omega$cm/K$^2$ at 15.5 T [8,12], indicative for the decrease of the quasiparticle-quasiparticle scattering cross section upon tuning the system away from the QCP. Similar behavior has been observed in the other NFL systems as well [23-25]. In YbRh$_2$Si$_2$, where weak AF order below $T_N=70$mK is suppressed at a critical magnetic field of $B_c = 0.06$ T (applied in the easy magnetic plane) a $1/(B-B_c)$ divergence of $A(B)$ has been observed very recently [26]. A more detailed analysis for UBe$_{13}$ is, however, hampered by the SC phase and the large cross-over field $B_0 \approx 14$ T. As for the other NFL systems, in UBe$_{13}$ the slope of the isothermal magnetoresistance changes from d$\rho(B)/$d$B<0$ at $B<B_0$ to d$\rho(B)/$d$B>0$ at $B>B_0$ [12]. To summarize, UBe$_{13}$ shows striking similarities to other NFL systems with well established AF quantum-critical points.

### 4. *B-T* phase diagram

As shown in the previous sections, the normal-state behavior of UBe$_{13}$ is compatible with an AF QCP covered by the SC phase. Measurements of the thermal expansion coefficient $\alpha(T)$ on U$_{1-x}$Th$_x$Be$_{13}$ have revealed, at $B=0$, a line of anomalies $T_L(x)$ in the $T$-$x$ diagram for $x<x_{c1}$. $T_L(x)$ marks the precursor of the lower of the two phase transitions at $T_{c2}$, found for $0.019 < x<0.046$ [9]. For pure UBe$_{13}$ a negative peak in $\alpha(T)$, clearly separated from the SC phase transition, has been attributed to short-range AF correlations below $T_L \approx 0.7$ K. An additional contribution that develops below $T_L$ is observed in the specific heat as well [10]. This becomes visible upon comparing the measured $C/T$ data with the $T$ dependence expected for the SC state (see solid line in Figure 1). The magnetic field dependence $T_L(B)$ as determined from thermal expansion and specific heat is plotted in the $B$-$T$ phase diagram, cf. Figure 4. Most importantly, $T_L$ vanishes around 4.5 T, i.e. almost the same field above which the universal NFL behavior in $\rho(T)$ is observed. Thus, one may speculate that the pronounced NFL effects observed in the normal state properties are related to a field-induced QCP, $T_L \rightarrow 0$ at about 4.5 T. It is interesting to note, that a clear enhancement of the slope d$T_c(B)/$d$B$ is observed near 4 T. Whether this stabilization of the SC state might be related to this "short-range AF" QCP remains unclear.

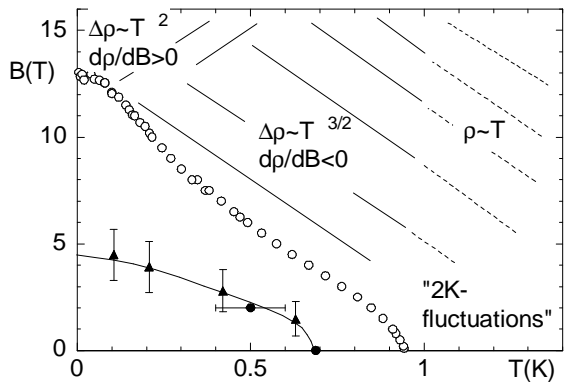

Fig. 4: *B-T* phase diagram for UBe$_{13}$. Open circles indicate SC phase boundary, defined by $\rho=0$, as derived from both *T*- and *B*-dependent measurements of the electrical resistivity [12]. Solid symbols indicate "line of anomalies" $B^*(T)$ obtained from temperature dependent thermal expansion (circles) and magnetic field dependent specific heat (triangles) measurements, respectively [9]. Existence ranges for different power-law behaviors in $\rho(T)$ as well as different signs of isothermal magnetoresistance slope $d\rho(B)/dB$ as discussed in text.

## 5. Conclusion

We have discussed the low-$T$ behavior of the HF metal UBe$_{13}$ for which an unconventional (yet not fully identified) SC ground state forms out an also unconventional normal state. The latter can be studied only in relatively large applied magnetic fields at lower temperatures, where it shows striking similarities with the normal state of other NFL superconductors, e.g. "S-type" CeCu$_2$Si$_2$[23], CeCoIn$_5$[24] and CeNi$_2$Ge$_2$[25]. At $B=0$, thermal expansion studies have revealed an anomaly at $T_L \approx 0.7$ K, that appears, due to its magnetic field dependence and its negative sign, indicative of the freezing out of AF short-range correlations [9]. We propose that the low-lying 3D AF spin fluctuations responsible for the NFL properties in the normal state of UBe$_{13}$ are those associated with the field-induced suppression of the "$T_L$-anomaly" at $B \approx 4.5$T.

## 6. Acknowledgements

Work at Florida was supported by the US Department of Energy, contract No. DE-FG05-86ER45268.

## References

1. J.G. Bednorz and K.A. Müller, Z. Phys. **B64**, 189 (1986).
2. F. Steglich et al., Phys. Rev. Lett. **43**, 1892 (1979).
3. N.D. Mathur et al., Nature **394**, 39 (1998).
4. H. Hegger et al., Phys. Rev. Lett. **84**, 4986 (2000).
5. H.R. Ott, et al., Phys. Rev. Lett. **50**, 1595 (1983).
6. D.L. Cox, Phys. Rev. Lett. **59**, 1240 (1987).
7. H.R. Ott et al., Phys. Rev. **B31**, 1651 (1985).
8. F. Steglich et al., Z. Phys. **B103**, 235 (1997).
9. F. Kromer et al., Phys. Rev. Lett. **81**, 4476 (1998).
10. R. Helfrich, Dissertation, TU Darmstadt (1997), unpublished.
11. C. Petrovic et al., J. Phys.: Condens. Matt. **13**, L337 (2001).
12. P. Gegenwart, Dissertation, TU Darmstadt (1998), unpublished.
13. D.L. Cox and M. Jarrell, J. Phys. Condens. Matt. **8**, 9825 (1996).
14. R. Felten et al., Europhys. Lett. **2**, 323 (1986).
15. S.L. Cooper et al., Phys. Rev. **B35**, 2615 (1987).
16. A.P. Ramirez et al., Phys. Rev. Lett. **73**, 3018 (1994).
17. E.A. Knetsch et al., Physica **B186-188**, 251 (1993).
18. M. Lang et al., Physica **B259-261**, 608 (1999).
19. See, for example, Y. Iye, in: D.M. Ginsberg (Ed.), Physical Properties of High-temperature Superconductors III, ch. 4, World Scientific, Singapore, 1992.
20. A.J. Millis, Phys. Rev. **B48**, 7183 (1993).
21. T. Moriya and T. Takimoto, J. Phys. Soc. Jpn. **64**, 960 (1995).
22. A. Rosch, Phys. Rev. Lett. **82**, 4280 (1999).
23. P. Gegenwart et al., Phys. Rev. Lett. **81**, 1501 (1998).
24. J. Paglione et al., cond-mat/0212502.
25. P. Gegenwart et al., Phys. Rev. Lett. **82**, 1293 (1999).
26. P. Gegenwart et al., Phys. Rev. Lett. **89**, 056402 (2002).